\documentclass[aps,prd,twocolumn,eqsecnum,showpacs,amsmath,nofootinbib]{revtex4-1}

\usepackage{color,graphicx}
\usepackage{amsfonts}
\usepackage{amssymb}
\begin{document}

\title{
Collisional Penrose Process in Rotating Wormhole Spacetime
}

\author{Naoki Tsukamoto} 

\author{Cosimo Bambi} 
\email[Corresponding author: ]{bambi@fudan.edu.cn}

\affiliation{
Center for Field Theory and Particle Physics and Department of Physics, Fudan University, 200433 Shanghai, China 
}
\date{\today}

\begin{abstract}
In a collisional Penrose process, two particles coming from the asymptotically flat region collide in the ergosphere of a compact object. The collision produces two new particles, one with positive energy and one with negative energy. When the particle with positive energy escapes to infinity, the process extracts energy from the compact object. In this paper, we study the collisional Penrose process in a rotating wormhole spacetime. We consider the simple case of a head-on collision at the throat of a Teo wormhole. We find that the process of energy extraction from a Teo wormhole can be substantially more efficient than the collisional Penrose process in the Kerr black hole spacetime.
\end{abstract}



\maketitle


\section{Introduction}

A rotating compact object can generate an ergoregion, namely a region of the spacetime in which any particle is forced to move along the direction of the rotation of the compact object. The interesting property of the ergoregion is that it allows the extraction of energy from the compact object. The simplest example is the Penrose process~\cite{pen71}. A particle coming from infinity enters the ergoregion of a black hole and decays into two particles, one with positive energy and one with negative energy. While the particle with negative energy is swallowed by the black hole, the particle with positive energy escapes to infinity and, since its energy is higher than that of the initial particle, the whole process has extracted energy from the black hole.

A variant of this mechanism is the collisional Penrose process, in which the initial state is represented by two particles coming from infinity that collide in the ergoregion of a black hole~\cite{Piran_Shaham_Katz_1975}. The product of the collision is a particle with positive energy that escapes to infinity and one with negative energy that crosses the event horizon. Once again, the whole process allows the extraction of rotational energy from the black hole. Recently, some authors have restudied the collisional Penrose process, both in the Kerr black hole spacetime~\cite{Bejger:2012yb,Schnittman:2014zsa} and in other stationary and axisymmetric spacetimes~\cite{Zaslavskii:2012yp}. Multi-Scattering scenarios in the Kerr spacetime have been investigated in~\cite{Grib:2010bs,Berti:2014lva}. The collisional Penrose process can even occur in a Reissner-Nordstr\"om black hole spacetime~\cite{Zaslavskii:2010aw}, since the latter has a generalized ergoregion~\cite{Denardo_Ruffini_1973}.

The aim of this paper is to study the collisional Penrose process in a rotating wormhole spacetime. Indeed, Penrose processes do not require the presence of an event horizon, but just of an ergoregion, and therefore they should be possible even in the case of rotating wormhole spacetimes. Wormholes are topologically non-trivial structures of the spacetime~\cite{Visser_1995}. They can be cast in general relativity as non-vacuum solutions. However, stationary and axisymmetric wormholes in general relativity necessarily require matter violating the null energy condition~\cite{Morris_Thorne_1988,Teo:1998dp}. The Ellis wormhole is the simplest wormhole solution in the Morris-Thorne class~\cite{Morris_Thorne_1988}. It was found in 1973 in Ref.~\cite{Ellis_1973} as a solution of the Einstein equations in the presence of a phantom scalar field and, independently, in Ref.~\cite{Bronnikov_1973} as a solution in a scalar-tensor theory of gravity. The Ellis wormhole metric can be obtained from different kinds of exotic matter~\cite{Das:2005un,Shatskiy:2008us}. The spacetime stability depends on the wormhole metric, the matter content as well as the gravity theory, and the stability of the Ellis wormhole has been extensively discussed in the literature~\cite{Shinkai_Hayward_2002}. For instance, the Ellis wormhole spacetime is linearly stable under spherical and axial perturbations~\cite{Bronnikov:2013coa} in the case where the matter content is a perfect fluid with negative density and a source-free radial electric or magnetic field~\cite{Shatskiy:2008us,Novikov:2012uj}. Lastly, we note there are also several papers on the possible observational signature of astrophysical wormholes and how they could be distinguished from black holes~\cite{observations}.

In the study of the collisional Penrose process, we need that the spacetime has an ergoregion and therefore that the wormhole is rotating. Unfortunately, analytical rotating wormhole solutions are not known at the moment. Only very recently, a numerical solution of a 4-dimensional rotating wormhole supported by a phantom scalar field has been obtained in Ref.~\cite{Kleihaus:2014dla}. For the sake of simplicity, in this paper we consider the Teo wormhole spacetime~\cite{Teo:1998dp}, even if it is not a complete solution with a known matter distribution, but just a wormhole metric. Particle collisions in the Teo wormhole have been investigated in~\cite{Tsukamoto:2014swa} to study the instability of a background spacetime against a particle collision with a high center-of-mass energy~\cite{Kimura:2010qy}. Here we study the simple case of a head-on collision at the throat of a Teo wormhole. As we expected, the presence of the wormhole ergoregion allows for the extraction of the rotational energy of the wormhole, namely the collisional Penrose process is possible and it is not a peculiarity of black hole spacetimes. Our result implies that the process of energy extraction can be much more efficient than that in the Kerr spacetime, and this is possible because we can create a head-on collision with the two particles coming from the two asymptotically flat regions. In the case of black holes, this is not possible simply because there is an event horizon and both the particles must come from the same flat region.

The paper is organized as follows. In Section~\ref{s-2}, we review the motion of a particle in the Teo wormhole spacetime. In Sections~\ref{s-3} and \ref{s-4}, we study the collisional Penrose process in the Teo wormhole spacetime. Summary and conclusions are reported in Section \ref{s-5}. Throughout the paper, we use geometrical units in which $c = G_{\rm N} = 1$.

\section{Teo Wormhole \label{s-2}}

The Teo wormhole spacetime is a simple stationary and axisymmetric wormhole spacetime. This section briefly reviews the motion of a test-particle in this spacetime.

\subsection{Line Element}

The line element of the Teo wormhole spacetime with the time translational Killing vector $t^{\mu}\partial_{\mu}=\partial_{t}$ and the axial Killing vector $\phi^{\mu}\partial_{\mu}=\partial_{\phi}$ is
\begin{eqnarray}\label{eq:Metric}
\hspace{-0.5cm}ds^{2}
&=&-N^{2}(r,\theta)dt^{2} +\frac{1}{1-\frac{b}{r}}dr^{2} \nonumber\\
&&+r^{2}K^{2}(r,\theta) \left[ d\theta^{2}+\sin^{2}\theta (d\phi-\omega(r) dt)^{2} \right], 
\end{eqnarray}
where $t, r, \theta$ and $\phi$ are spherical polar coordinates. They are defined in the range $-\infty\leq t\leq \infty$, $b\leq r \leq \infty$, $0\leq \theta\leq \pi$, and $0\leq \phi\leq 2\pi$, where $b$ is a positive constant and the functions $N(r,\theta)$, $K(r,\theta)$ and $\omega(r)$ are given by
\begin{eqnarray}\label{eq:N}
N(r,\theta)
&=&K(r,\theta)
=1+\frac{16a^{2}d\cos^{2}\theta}{r},\\
\omega(r)
&=&\frac{2a}{r^{3}} .
\end{eqnarray}
Here $a$ is the wormhole spin parameter and $d$ is a positive constant~\footnote{We note that we have introduced the parameter $d$ to tune the dimensions of the second term on the right-hand side in Eq.~(\ref{eq:N}). If we set $b=d=1$, we obtain the same line element as in Ref.~\cite{Teo:1998dp}. }. In what follows, we assume that the angular momentum of the wormhole $a$ is non-negative without loss of generality.

The wormhole has a throat at $r=b$ and the flare-out condition~\cite{Morris_Thorne_1988} is satisfied there:
\begin{eqnarray}
\frac{1}{2b^{2}}\left( b-\frac{db}{dr}r \right)=\frac{1}{2b}>0.
\end{eqnarray}
The shape of the wormhole throat is like a peanut shell (see Fig.~1 in~\cite{Teo:1998dp}). When $b^{2}/2 \leq a$, the wormhole has an ergoregion in $b^{2} \leq r^{2} \leq 2a \sin \theta $. In the ergoregion, a particle can have a negative conserved energy. In such a case, the particle is trapped inside the ergoregion. We note that the ergoregion can never extend to the rotational axes 
($\theta=0$, $\pi$) and that an event horizon never forms there, even when the angular momentum of the wormhole assumes a very high value $b^{2}\ll a$.

\subsection{Hamilton-Jacobi Equation}

The motion of a test-particle with rest mass $m$ can be obtained by solving the Hamilton-Jacobi equation:
\begin{eqnarray}
\frac{\partial S}{\partial \lambda} +H=0,
\end{eqnarray}
where $S=S(\lambda, x^{\nu})$ is the action, which is a function of the coordinate $x^{\nu}$ and of the affine parameter $\lambda \equiv \tau/m$, $\tau$ is the proper time, and $H$ is the Hamiltonian of the particle. The latter is defined as $H\equiv g_{\mu\nu}p^{\mu}p^{\nu}/2$, where $p^{\mu}=dx^{\mu}/d\lambda$ is the conjugate momentum and
\begin{eqnarray}
p_{\mu}=\frac{\partial S}{\partial x^{\mu}}.
\end{eqnarray}
Since $t$ and $\phi$ are the cyclic coordinates, the action $S$ can be cast in the following form
\begin{eqnarray}
S=\frac{1}{2}m^{2}\lambda -Et +L\phi +S_{r}(r) +S_{\theta}(\theta),
\end{eqnarray}
where $E\equiv -p_{\mu}t^{\mu}=-p_{t}$ and $L\equiv p_{\mu}\phi^{\mu}=p_{\phi}$ are, respectively, the conserved energy and the conserved angular momentum of the test-particle.

The Hamilton-Jacobi equation can be written as
\begin{eqnarray}\label{eq:Hamilton-Jacobi}
&&\left( r+16a^{2}d\cos^{2}\theta \right)^{2} m^{2}
-r^{2}\left( E-\frac{2aL}{r^{3}} \right)^{2} \nonumber\\
&&+\frac{L^{2}}{\sin^{2}\theta}
+\left( r+16a^{2}d\cos^{2}\theta \right)^{2} \left( 1-\frac{b}{r} \right) 
\left( \frac{dS_{r}}{dr} \right)^{2} \nonumber\\
&&+\left( \frac{dS_{\theta}}{d\theta} \right)^{2}
=0.
\end{eqnarray}
This equation is not separable in $r$ and $\theta$ because of the terms with the factor $r\cos^{2}\theta$. We note that the Hamilton-Jacobi equation is separable in the case $d=0$~\cite{Mori_2014}, while here we want to consider the case $d>0$. The conjugate momenta $p^{\mu}=dx^{\mu}/d\lambda$ are
\begin{eqnarray}
&&p^{t}=\frac{E-\omega(r) L}{N^{2}(r,\theta)},\\
&&p^{r}=\left( 1-\frac{b}{r} \right) \frac{dS_{r}}{dr},\\
&&p^{\theta}=\frac{1}{r^{2}K^{2}(r,\theta)}\frac{dS_{\theta}}{d\theta},\\
&&p^{\phi}=\frac{\omega(r) (E-\omega(r) L)}{N^{2}(r,\theta)}+\frac{L}{r^{2}K^{2}(r,\theta)\sin^{2}\theta}.
\end{eqnarray}
The forward-in-time condition $dt/d\lambda=p^{t} \geq 0$ must be satisfied along the geodesic and therefore
\begin{eqnarray}\label{Definition_mathcal_E}
\mathcal{E}(r) \equiv E-\omega(r) L \geq 0 
\end{eqnarray}
along the geodesic.

\subsection{Effective Potential of a Particle on the Equatorial Plane}

Let us now consider a particle moving on the equatorial plane $\theta=\pi/2$. If we introduce the new radial coordinate $\rho$ defined by
\begin{eqnarray}\label{eq:Definition_rho}
\frac{d\rho}{dr}\equiv \pm \left( 1-\frac{b}{r} \right)^{-\frac{1}{2}}
\end{eqnarray}
and in the range $-\infty < \rho <\infty$, we can integrate Eq.~(\ref{eq:Definition_rho}) and we find
\begin{eqnarray}
\hspace{-0.5cm}
\rho=\pm \left[ \sqrt{r(r-b)}+b\log{\left( \sqrt{\frac{r}{b}}+\sqrt{\frac{r}{b}-1} \right)} \right],
\end{eqnarray}
where we have set the wormhole throat at $\rho=0$ with the choice of the integration constant. The line element on the equatorial plane with the new radial coordinate $\rho$ is
\begin{eqnarray}
ds^{2}=-dt^{2} +d\rho^{2} +r^{2}(\rho) \left( d\phi-\omega(\rho)dt  \right)^{2}.
\end{eqnarray}
Imposing $\theta=\pi/2$, $dS_{\theta}/d\theta=0$, and
\begin{eqnarray}
\left( \frac{d\rho}{d\lambda}\right)^{2}=\left( 1-\frac{b}{r} \right) \left( \frac{dS_{r}}{dr} \right)^{2}
\end{eqnarray}
in the Hamilton-Jacobi equation in~(\ref{eq:Hamilton-Jacobi}), we obtain the energy equation for a particle moving on the equatorial plane
\begin{eqnarray}\label{eq:energy_eq}
\frac{1}{2}\left( \frac{d\rho}{d\lambda} \right)^{2} +V_{\mathrm{eff}}(\rho)=0,
\end{eqnarray}
where the effective potential $V_{\mathrm{eff}}(\rho)$ is 
\begin{eqnarray}\label{eq:Effective_potential}
V_{\mathrm{eff}}(\rho)\equiv \frac{1}{2}\left( m^{2}-\mathcal{E}^{2}(\rho) +\frac{L^{2}}{r^{2}(\rho)} \right).
\end{eqnarray}
From the energy equation~(\ref{eq:energy_eq}) and $p^{\rho}(\rho)=d\rho/d\lambda$, we can rewrite the conjugate momentum of $\rho$ as
\begin{eqnarray}\label{eq:p_rho_VS_potential}
p^{\rho}(\rho)=\sigma_{\rho}\sqrt{-2V_{\mathrm{eff}}(\rho)},
\end{eqnarray}
where $\sigma_{\rho}\equiv p^{\rho}(\rho)/ \left| p^{\rho}(\rho) \right| =\pm 1$. In the region $\rho \geq 0$, a particle approaches the throat if $\sigma_{\rho}=-1$ and moves away from the throat if $\sigma_{\rho}=1$. On the contrary, in the region $\rho < 0$, a particle approaches the throat if $\sigma_{\rho}=1$ and moves away from the throat if $\sigma_{\rho}=-1$. A particle is constrained to move in the region $V_{\mathrm{eff}}(\rho)\leq 0$. If $E^{2}\geqq m^{2}$, the particle can be at infinity because
\begin{eqnarray}
\lim_{\rho \rightarrow \pm \infty} V_{\mathrm{eff}}(\rho)= \frac{1}{2} \left( m^{2}-E^{2} \right).
\end{eqnarray}
The derivative of the effective potential with respect to $\rho$ is
\begin{eqnarray}\label{eq:dV_drho}
V_{\mathrm{eff}}'(\rho)
=\mp \sqrt{1-\frac{b}{r(\rho)}} \left[ \frac{6aL \mathcal{E}(\rho)}{r^{4}(\rho)} +\frac{L^{2}}{r^{3}(\rho)} \right], 
\end{eqnarray}
where the upper (lower) sign holds in the region $\rho > 0$ ($\rho <0$). The second derivative of the effective potential with respect to $\rho$ is given by
\begin{widetext}
\begin{eqnarray}
V_{\mathrm{eff}}''(\rho)
=-\frac{bL}{2r^{6}(\rho)} \left( 6a\mathcal{E}(\rho)+Lr(\rho) \right) +  \left( 1-\frac{b}{r(\rho)} \right) \frac{3L \left( 8aEr^{3}(\rho)-24a^{2}L +Lr^{4}(\rho) \right)}{r^{8}(\rho)} . 
\end{eqnarray}
\end{widetext}

\section{Conservation Laws \label{s-3}}

In what follows, we assume a Teo wormhole spacetime with $\sqrt{2a}\geq b$. In our collisional Penrose process, we have two particles, say~(1) and~(2), that collide at the wormhole throat $\rho=0$. The collision creates two new particles, (3) and~(4). To simplify our discussion, we assume  
\begin{eqnarray}\label{eq:Assumption_E_m_12}
E_{(1)}=E_{(2)} \geq m_{(1)}=m_{(2)}\geq 0
\end{eqnarray}
and 
\begin{eqnarray}\label{eq:Assumption_L_12}
L_{(1)}=L_{(2)},
\end{eqnarray}
where the subscripts $(1)$, $(2)$, $(3)$ and $(4)$ refer to the particles~$(1)$, $(2)$, $(3)$ and $(4)$, respectively. Under the assumption~(\ref{eq:Assumption_E_m_12}), the particles~$(1)$ and $(2)$ can exist at infinity. Indeed, $V_{\mathrm{eff} (1)}(\pm \infty) = V_{\mathrm{eff} (2)}(\pm \infty) =\left( m_{(1)}^{2}-E_{(1)}^{2} \right)/2 \leq 0$. From the assumptions~(\ref{eq:Assumption_E_m_12}) and (\ref{eq:Assumption_L_12}) follow that both $\mathcal{E}(\rho)$ and the effective potential~$V_{\mathrm{eff}}(\rho)$, defined respectively in Eqs.~(\ref{Definition_mathcal_E}) and (\ref{eq:Effective_potential}), are the same for particle~$(1)$ and particle~$(2)$, namely
\begin{eqnarray}
\mathcal{E}_{(1)}(\rho)=\mathcal{E}_{(2)}(\rho), \, \,\, 
V_{\mathrm{eff}(1)}(\rho)=V_{\mathrm{eff}(2)}(\rho).
\end{eqnarray}

Let us assume that the particle~$(1)$ is approaching the throat from the region $\rho> 0$, while the particle~$(2)$ is approaching the throat from the region $\rho <0$. With this set-up we have $\sigma_{\rho(1)} = -1$ and $\sigma_{\rho(2)} = 1$, and the $\rho$ component of the total 4-momentum of the particles~$(1)$ and $(2)$ vanishes at the collisional point
\begin{eqnarray}\label{eq:Four_momentum_rho}
p^{\rho}_{(1)}(0)+p^{\rho}_{(2)}(0)=0.
\end{eqnarray}
Let us also assume that the particles~(3) and (4) have the same non-negative mass, that the conserved energy of the particle~(3) is larger than or equal to the mass, while the particle~(4) has non-positive energy
\begin{eqnarray}\label{eq:Assumption_m_34}
&&E_{(3)}\geq  m_{(3)}=m_{(4)}\geq 0, \\
\label{eq:Assumption_negative_energy_4}
&&E_{(4)}\leq 0. 
\end{eqnarray}
The particle~$(4)$ is thus bound in the ergoregion~$-\rho_{\rm ergo} \leq \rho \leq \rho_{\rm ergo}$, 
where $\rho_{\rm ergo}$ is defined by
\begin{eqnarray}
\rho_{\rm ergo} \equiv \sqrt{\sqrt{2a}(\sqrt{2a}-b)}
+b\log{\left( \sqrt{\frac{\sqrt{2a}}{b}}+\sqrt{\frac{\sqrt{2a}}{b}-1} \right)}.\nonumber\\
\end{eqnarray}

\subsection{Center-of-Mass Energy and Mass Upper Bound}

The center-of-mass energy $E_{\mathrm{CM}}(\rho)$ of the particles~$(1)$ and $(2)$ at the throat on the equatorial plane in the Teo wormhole spacetime is given by~\cite{Tsukamoto:2014swa}
\begin{eqnarray}\label{eq:CM_energy}
E_{\mathrm{CM}}^{2}(0)
&\equiv&-\left(p^{\mu}_{(1)}(0)+p^{\mu}_{(2)}(0) \right) \left(p_{(1)\mu}(0)+p_{(2)\mu}(0) \right) \nonumber\\
&=&4\mathcal{E}_{(1)}^{2}(0)-\frac{4L^{2}_{(1)}}{b^{2}}.
\end{eqnarray}
From the local conservation of energy, we have
\begin{eqnarray}\label{eq:Total_mass_VS_CM_energy}
m_{(3)}+m_{(4)}\leq  E_{\mathrm{CM}}(0).
\end{eqnarray} 
From $m_{(3)}=m_{(4)}$ and Eqs.~(\ref{eq:CM_energy}) and (\ref{eq:Total_mass_VS_CM_energy}), we find an upper bound for the rest mass of the particle~$(3)$ created in the collision
\begin{eqnarray}\label{eq:Upper_bound_Mass}
m_{(3)}^{2}
\leq \mathcal{E}_{(1)}^{2}(0)-\frac{L^{2}_{(1)}}{b^{2}}.
\end{eqnarray}

\subsection{Conservation of the 4-Momentum}

From the conservation equation of the total 4-momentum, at the collision point $\rho=0$ before and after the particle collision we have
\begin{eqnarray}\label{eq:Conservation_four_momentum}
p^{\mu}_{(1)}(0)+p^{\mu}_{(2)}(0)=p^{\mu}_{(3)}(0)+p^{\mu}_{(4)}(0).
\end{eqnarray}
From the $t$, $\phi$ and $\rho$ components of the conservation equation of the total 4-momentum~(\ref{eq:Conservation_four_momentum}), we obtain 
\begin{eqnarray}\label{eq:Conservation_E}
&&E_{(1)}+E_{(2)}=E_{(3)}+E_{(4)}, \\
\label{eq:Conservation_L}
&&L_{(1)}+L_{(2)}=L_{(3)}+L_{(4)}, \\
\label{eq:Conservation_P_rho}
&&p^{\rho}_{(1)}(0)+p^{\rho}_{(2)}(0)=p^{\rho}_{(3)}(0)+p^{\rho}_{(4)}(0).
\end{eqnarray}

Under the assumptions~(\ref{eq:Assumption_E_m_12}), (\ref{eq:Assumption_L_12}) and (\ref{eq:Assumption_negative_energy_4}), using the conservation equations of the conserved energy~(\ref{eq:Conservation_E}) and of the conserved angular momentum~(\ref{eq:Conservation_L}),
we can write the conserved energy and the conserved angular momentum of the particle~$(4)$ as  
\begin{eqnarray}\label{eq:E_4}
&&E_{(4)}=2E_{(1)}-E_{(3)}\leq 0 , \\
\label{eq:L_4}
&&L_{(4)}=2L_{(1)}-L_{(3)}.
\end{eqnarray} 
From Eqs.~(\ref{eq:p_rho_VS_potential}) and (\ref{eq:Four_momentum_rho}), Eq.~(\ref{eq:Conservation_P_rho}) can be written as
\begin{eqnarray}\label{eq:Local_conservation_rho_four_momenta}
\hspace{-0.5cm}
0=\sigma_{\rho (3)}\sqrt{-2V_{\mathrm{eff}(3)}(0)}+\sigma_{\rho (4)}\sqrt{-2V_{\mathrm{eff}(4)}(0)}.
\end{eqnarray}
We note that $\sigma_{\rho (3)}\sigma_{\rho (4)}=-1$ must be satisfied. Moreover, the effective potential~(\ref{eq:Effective_potential}) is even with respect to $\rho$, namely $V_{\mathrm{eff}}(-\rho)=V_{\mathrm{eff}}(\rho)$, and therefore if a particle produced in the collision is reflected by a potential barrier that particle is bound near the throat. Here we are interested in the case in which the particle~$(3)$ escapes to infinity ($\rho \rightarrow \infty$) and the particle~$(4)$ is trapped in the vicinity of the wormhole throat. We can also assume $\sigma_{\rho (3)}=1$ and $\sigma_{\rho (4)}=-1$. In what follows, we just consider the case
\begin{eqnarray}\label{eq:Potential_3_vs_4}
V_{\mathrm{eff}(3)}(0)=V_{\mathrm{eff}(4)}(0) \leq 0.
\end{eqnarray}
which satisfies Eq.~(\ref{eq:Local_conservation_rho_four_momenta}).

\section{Energy Extraction from a Teo Wormhole \label{s-4}}

We consider two scenarios of collision. The flat potential case, in which the initial particles have vanishing conserved angular momentum, and the deep potential case, in which the initial particles have negative conserved angular momentum.

\subsection{Case with Flat Potential}

Let us assume that the particles~(1) and (2) have zero conserved angular momentum, namely
\begin{eqnarray}\label{eq:Assumption_L_12_zero}
L_{(1)}=L_{(2)}=0.
\end{eqnarray}
In this case, the effective potentials of the particles~(1) and (2) are ``flat'' 
\begin{eqnarray}
V_{\mathrm{eff}(1)}(\rho)=V_{\mathrm{eff}(2)}(\rho)=\frac{1}{2}\left( m_{(1)}^{2}-E_{(1)}^{2} \right).
\end{eqnarray} 
From the assumption~(\ref{eq:Assumption_E_m_12}), the effective potentials of the particles~(1) and (2) are non-positive and the forward-in-time condition is satisfied everywhere, $\mathcal{E}_{(1)}(\rho)=\mathcal{E}_{(2)}(\rho)=E_{(1)}>0$. In this case, the particles~$(1)$ and $(2)$ are initially in the asymptotically flat regions and they can reach the throat, where they collide. The center-of-mass energy of the particles~(1) and (2) at the throat $\rho=0$ is given by
\begin{eqnarray}
E_{\mathrm{CM}}(0)=2E_{(1)}.
\end{eqnarray} 
This leads to an upper bound for the rest mass of the particle~(3) produced in the particle collision
\begin{eqnarray}
m_{(3)}\leq E_{(1)}.
\end{eqnarray} 
From Eq.~(\ref{eq:L_4}), the conserved angular momentum of the particle~(4) is given by
\begin{eqnarray}\label{eq:L_4_Flat}
L_{(4)}=-L_{(3)}.
\end{eqnarray} 
If we plug Eqs.~(\ref{eq:E_4}) and (\ref{eq:L_4_Flat}) into Eq.~(\ref{eq:Potential_3_vs_4}), we get
\begin{eqnarray}\label{eq:L_3_Flat}
L_{(3)}=\frac{b^{3}}{2a} \left( E_{(3)}-E_{(1)} \right).
\end{eqnarray}
From Eqs.~(\ref{eq:Assumption_E_m_12}), (\ref{eq:E_4}), (\ref{eq:L_4_Flat}) and (\ref{eq:L_3_Flat}), $L_{(3)}$ is positive and $L_{(4)}$ is negative. The inequality~(\ref{eq:Potential_3_vs_4}) is satisfied if and only if 
\begin{eqnarray}\label{eq:V4_zero}
E_{-}
\leq E_{(3)}
\leq E_{+}
\end{eqnarray}
is satisfied, where $E_{\pm}$ is defined as
\begin{eqnarray}\label{eq:Def_E+}
E_{\pm}\equiv E_{(1)} \pm\frac{2a}{b^{2}}\sqrt{E_{(1)}^{2}-m_{(3)}^{2}}.
\end{eqnarray}
The inequality (\ref{eq:V4_zero}) gives an upper bound of the energy extraction efficiency $\eta$
\begin{eqnarray}
\eta
\equiv \frac{E_{(3)}}{E_{(1)}+E_{(2)}}
\leq \eta_{+},
\end{eqnarray}
where $\eta_{+}$ is defined as
\begin{eqnarray}
\eta_{+}
\equiv \frac{E_{+}}{E_{(1)}+E_{(2)}}
=\frac{1}{2}+\frac{a}{b^{2}}\sqrt{1-\frac{m_{(3)}^{2}}{E_{(1)}^{2}}}.
\end{eqnarray}
From Eqs.~(\ref{eq:E_4}), (\ref{eq:V4_zero}) and (\ref{eq:Def_E+}), we obtain an upper bound for the rest mass of the particle~(3), namely
\begin{eqnarray}\label{eq:upper_bound_rest_mass}
m_{(3)}\leq \sqrt{1-\frac{b^{4}}{4a^{2}}}E_{(1)},
\end{eqnarray}
which is related to the energy extraction. Using Eqs.~(\ref{eq:L_4_Flat}) and (\ref{eq:L_3_Flat}), we obtain the function $\mathcal{E}(\rho)$ for the particles~$(3)$ and $(4)$
\begin{eqnarray}\label{eq:Forward-in-time_condition_3}
\hspace{-0.5cm}
&&\mathcal{E}_{(3)}(\rho)=E_{(1)}+\left( E_{(3)}-E_{(1)} \right)\left( 1-\frac{b^{3}}{r^{3}(\rho)} \right),\\
\hspace{-0.5cm}
&&\mathcal{E}_{(4)}(\rho)=E_{(1)}-\left( E_{(3)}-E_{(1)} \right)\left( 1-\frac{b^{3}}{r^{3}(\rho)} \right).
\end{eqnarray}
From Eq.~(\ref{eq:Forward-in-time_condition_3}), the forward-in-time condition for particle (3) is satisfied, since $\mathcal{E}_{(3)}(\rho)$ is positive everywhere. Since $L_{(3)}$ and $\mathcal{E}_{(3)}(\rho)$ are positive, in the region $\rho \geq 0$ we have that $V_{\mathrm{eff}(3)}'(\rho)$ is non-positive and the effective potential~$V_{\mathrm{eff}(3)}(\rho)$ is a monotonically decreasing function with respect to $\rho$. It follows that $V_{\mathrm{eff} (3)}(\rho)\leq 0$ everywhere, because $V_{\mathrm{eff} (3)}(0)\leq 0$. The particle~(3) created at the throat can thus escape to infinity $\rho \rightarrow \infty$.

Let us introduce $\rho_{A}$
\begin{eqnarray}
\rho_{A} 
\equiv b \sqrt{A(A-1)} +b\log \left( \sqrt{A} +\sqrt{A-1} \right), 
\end{eqnarray}
where $A$ is
\begin{eqnarray}\label{eq:Def_A}
A
\equiv\frac{1}{b} \left( \frac{2aL_{(4)}}{E_{(4)}} \right)^{\frac{1}{3}} 
=\left( \frac{E_{(3)}-E_{(1)}}{E_{(3)}-2E_{(1)}} \right)^{\frac{1}{3}}.
\end{eqnarray}
From Eqs.~(\ref{eq:Assumption_E_m_12}) and (\ref{eq:E_4}), $A$ is larger than unity. At $\rho=\pm \rho_{A}$, $r(\pm \rho_{A})=Ab$ and $\mathcal{E}_{(4)}(\pm \rho_{A})=0$. The effective potential for the particle~(4) at $\rho=\rho_{A}$ is thus
\begin{eqnarray}
V_{\mathrm{eff} (4)}\left( \rho_{A} \right) 
=\frac{1}{2} \left( m_{(4)}^{2}+\frac{L_{(4)}^{2}}{A^{2} b^{2}} \right) > 0.
\end{eqnarray}
Since the effective potential for the particle~(4) is non-positive at the throat $\rho=0$, it is positive at the point $\rho=\rho_{A}$, and it is a continuous function with respect to $\rho$, there is at least one turning point in the region~$0\leq \rho <\rho_{A}$. This implies that the particle~$(4)$ is trapped in the region~$-\rho_{t} \leq \rho \leq \rho_{t}$, where $\rho_{t}$ is the minimum turning point for the particle~$(4)$. We note that the forward-in-time condition $\mathcal{E}_{(4)}(\rho)\geq 0$ for the particle~$(4)$ is satisfied within the two turning points $-\rho_{t}\leq \rho \leq \rho_{t}$. Actually, the particle~(4) is trapped inside the ergoregion $-\rho_{\rm ergo} \leq \rho \leq \rho_{\rm ergo}$, as it can be seen in the following way. We consider the case in which the particle~(3) has the maximal conserved energy $E_{(3)}=E_{+}$, corresponding to the maximum energy extraction efficiency. The effective potential of the particle~(4) and its derivative with respect to $\rho$ vanish at the throat, $V_{\mathrm{eff} (4)}(0)=V_{\mathrm{eff} (4)}'(0)=0$. The second derivative of the effective potential of the particle~(4) with respect to $\rho$ at the throat is given by
\begin{eqnarray}
V_{\mathrm{eff} (4)}'' \left( 0 \right) 
=\frac{\sqrt{E_{(1)}^{2}-m_{(3)}^{2}}}{2b^{4}} \left( 6aE_{(1)}-b^{2}\sqrt{E_{(1)}^{2}-m_{(3)}^{2}} \right). \nonumber\\
\end{eqnarray}
From the inequality~(\ref{eq:upper_bound_rest_mass}) and $\sqrt{2a}\geq b$, $V_{\mathrm{eff} (4)}'' \left( 0 \right)$ is positive. Thus, the particle~(4) is trapped at the wormhole throat and it cannot exit the ergoregion. An example of effective potential of the particle~(4) is shown in Fig.~1.

\begin{figure}[t]
\begin{center}
\includegraphics[width=80mm]{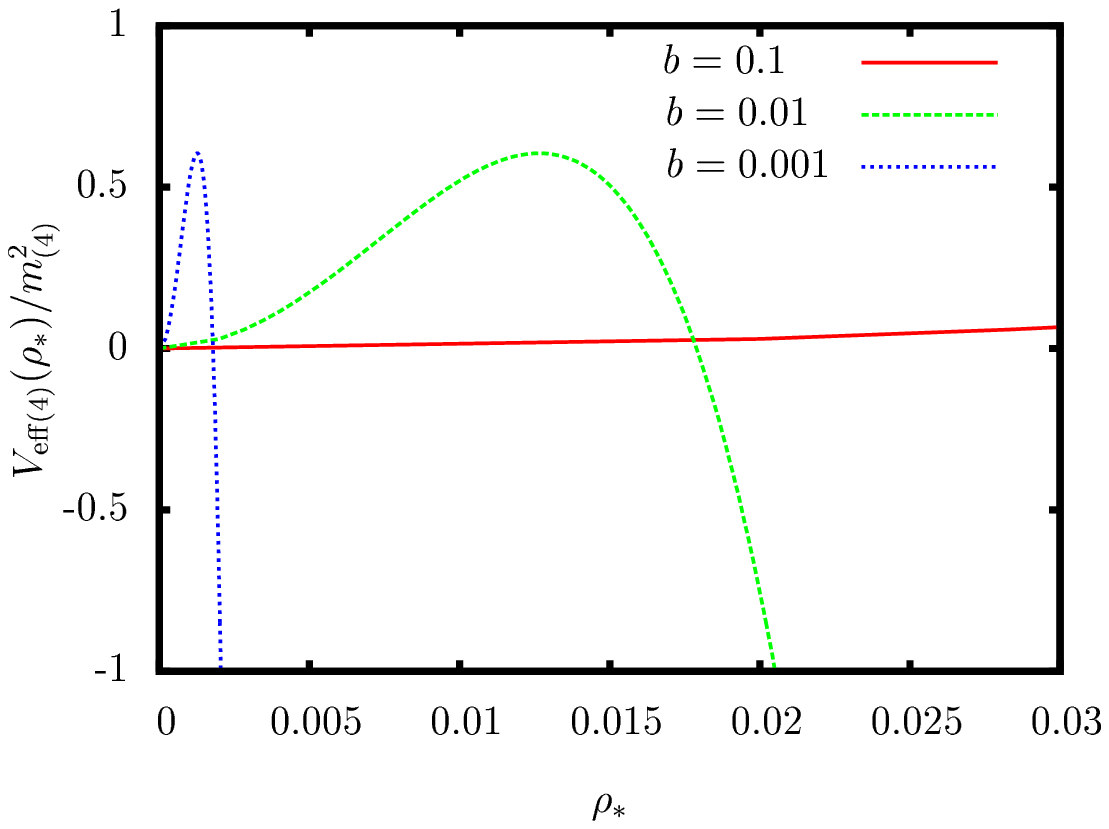}
\end{center}
\caption{Examples of $V_{\mathrm{eff}(4)}(\rho_{*})/m_{(4)}^{2}$, where $\rho_{*}\equiv \rho/b$. The red-solid, green-dashed, and blue-dotted curves denote the effective potentials of the particle~(4) when $b=0.1$, $b=0.01$ and $b=0.001$, respectively. Here $E_{(3)}=E_{+}$, $m_{(4)}=1$, $E_{(1)}=1.1$ and $a=1$. We note that both the effective potential of the particle~(4) and its derivative with respect to $\rho$ vanish at the wormhole throat, $V_{\mathrm{eff}(4)}(0)=V_{\mathrm{eff}(4)}'(0)=0$.}
\end{figure}

\subsection{Case with Deep Potential}

As a second scenario, we consider a case in which the two incident particles have negative conserved angular momentum $L_{(1)}=L_{(2)}<0$. We assume that the conserved angular momentum of the particle~(3) is
\begin{eqnarray}\label{Assume_forward_in_time_3}
L_{(3)}\leq \frac{b^{3}}{2a}E_{(3)}.
\end{eqnarray}
Under this assumption, the forward-in-time condition $\mathcal{E}_{(3)}(\rho)\geq 0$  for the particle~(3) is satisfied everywhere. We also assume that the particle~(4) satisfies the forward-in-time condition at the throat
\begin{eqnarray}\label{Assume_forward_in_time_4}
\mathcal{E}_{(4)}(0)\geq 0.
\end{eqnarray} 
The inequalities~(\ref{Assume_forward_in_time_3}) and (\ref{Assume_forward_in_time_4}) give
\begin{eqnarray}\label{Assume_forward_in_time_5}
\frac{b^{3}}{2a}\left( E_{(3)}-2E_{(1)} \right) +2L_{(1)} \leq  L_{(3)}\leq \frac{b^{3}}{2a}E_{(3)}.
\end{eqnarray}
In this case, the particles~(1) and (2) satisfy the forward-in-time condition everywhere
\begin{eqnarray}
\mathcal{E}_{(1)}(\rho)
=\mathcal{E}_{(2)}(\rho)
=E_{(1)}-\frac{2aL_{(1)}}{r^{3}(\rho)}
\geq 0.
\end{eqnarray}
From Eqs.~(\ref{eq:E_4}), (\ref{eq:L_4}) and (\ref{eq:Potential_3_vs_4}), we obtain
\begin{eqnarray}\label{eq:L_3_deep}
L_{(3)}=\frac{\mathcal{E}_{(1)}(0) \left( E_{(3)}-\mathcal{E}_{(1)}(0) \right) +\frac{L_{(1)}^{2}}{b^{2}}}{\frac{2aE_{(1)}}{b^{3}}+\frac{L_{(1)}}{b^{2}}\left( 1-\frac{4a^{2}}{b^{4}} \right)}.
\end{eqnarray}
Fig.~2 shows some examples of effective potential for the particle~(1) in which the particle can reach the wormhole throat from the flat region $\rho>0$. In this case, the effective potentials of the particles~(1) and (2) are ``deep'' and the center-of-mass energy is very large if $a/b^{2}\gg 1$.

\begin{figure}[b]
\begin{center}
\includegraphics[width=80mm]{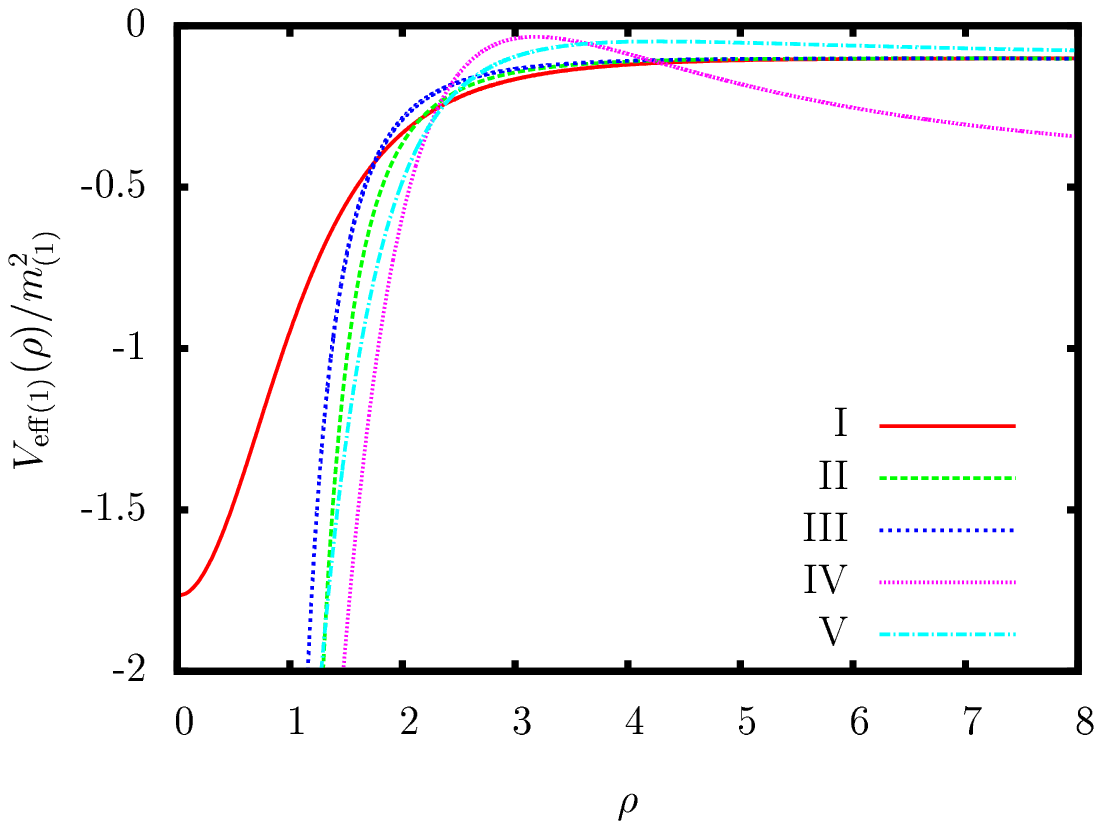}
\end{center}
\caption{Examples of $V_{\mathrm{eff}(1)}(\rho)/m_{(1)}^{2}$. The red-solid, green-dashed, blue-dot-spaced, magenta-dotted, and cyan-dashed-dotted curves denote the effective potentials for the motion with respect to the radial direction $\rho$ of the particle~(1) with, respectively, 
the following set of parameters: 
I~($b=2^{1/4}$, $E_{(1)}=1.1$, $L_{(1)}=-1$),
II~($b=0.1$, $E_{(1)}=1.1$, $L_{(1)}=-1$),
III~($b=0.001$, $E_{(1)}=1.1$, $L_{(1)}=-1$),
IV~($b=2^{1/4}$ $E_{(1)}=1.4$, $L_{(1)}=-4$) 
and
V~($b=1$, $E_{(1)}=1.1$, $L_{(1)}=-2$).
Here $m_{(1)}=1$ and $a=1$. }
\end{figure}

The inequality~(\ref{eq:Potential_3_vs_4}) is satisfied if and only if 
\begin{eqnarray}\label{eq:E_pm}
E_{-}\leq E_{(3)}\leq E_{+},
\end{eqnarray}
where $E_{\pm}$ is defined as
\begin{eqnarray}\label{eq:Def_E_pm}
\hspace{-0.8cm} E_{\pm}
\equiv 
E_{(1)} \pm \sqrt{B} \left[ \frac{2aE_{(1)}}{b^{2}}+\frac{L_{(1)}}{b}\left( 1-\frac{4a^{2}}{b^{4}} \right) \right], 
\end{eqnarray}
and $B$ is
\begin{eqnarray}
B\equiv 1-\frac{m_{(3)}^{2}}{\mathcal{E}_{(1)}^{2}(0)-\frac{L_{(1)}^{2}}{b^{2}}}.
\end{eqnarray}
From Eqs.~(\ref{eq:Assumption_m_34}) and (\ref{eq:Upper_bound_Mass}), $B$ is non-negative and it is not larger than 1, $0\leq B\leq 1$. From the inequality~(\ref{eq:E_pm}), we obtain the maximal energy extraction efficiency given by 
\begin{eqnarray}
\eta_{+}
=\frac{1}{2} +\frac{\sqrt{B}}{2} \left[ \frac{2a}{b^{2}}+\frac{L_{(1)}}{bE_{(1)}}\left( 1-\frac{4a^{2}}{b^{4}} \right) \right].
\end{eqnarray}
For $B=1/2$, the maximal energy extraction efficiency with the set of parameters adopted in Fig.~2 is $\eta_{+}\sim 1.3$ (case I), $1.3 \times 10^{5}$ (II), $1.3 \times 10^{15}$ (III), $1.8$ (IV), and $3.1$ (V). The efficiency can thus be huge for suitable choices of the model parameters. From the assumptions that the particle~(4) satisfies the forward-in-time condition at the throat~(\ref{Assume_forward_in_time_4}) and that the conserved energy of the particle~(4) is negative~(\ref{eq:Assumption_negative_energy_4}), we obtain $L_{(4)}\leq 0$ and $1 \leq A$, where $A$ is
\begin{eqnarray}\label{eq:Def_A_again}
A \equiv\frac{1}{b} \left( \frac{2aL_{(4)}}{E_{(4)}} \right)^{\frac{1}{3}}.
\end{eqnarray}
The particle~(4) turns out to be bound in the region $-\rho_{t} \leq \rho \leq \rho_{t}$, where $\rho_{t}$ is the minimum and positive turning point for the particle~(4) and the forward-in-time condition for the particle~(4) is satisfied there in the same way as the case of the flat effective potential case.

For simplicity, in what follows we concentrate on the case $E_{(3)}=E_{+}$, $a/b^{2} \geq \sqrt{2}/2$ and $B=1/2$. From Eqs.~(\ref{eq:L_3_deep}) and (\ref{eq:Def_E_pm}), $L_{(3)}$ is positive. The particle~(3) with the rest mass $m_{(3)}=E_{\mathrm{CM}}(0)/(2\sqrt{2})$ can thus escape to infinity ($\rho \rightarrow +\infty$). Such a conclusion can be derived proceeding in the same way as we did for the flat potential case. It is also easy to show that the condition~(\ref{Assume_forward_in_time_5}), or, equivalently, the assumptions~(\ref{Assume_forward_in_time_3}) and (\ref{Assume_forward_in_time_4}), are satisfied in this case.

Finally, we can check that the particle~(4) with the negative conserved energy $E_{(4)}\leq -2E_{(1)}$ is trapped in the ergoregion. Both the effective potential and its derivative with respect to $\rho$ vanish at the throat, $V_{\mathrm{eff}(4)}(0)=V_{\mathrm{eff}(4)}'(0)=0$. The second derivative of the effective potential with respect to $\rho$ at the throat is
\begin{eqnarray}
V_{\mathrm{eff}(4)}''(0)=-\frac{L_{(4)}f}{2b^{3}g},
\end{eqnarray}
where
\begin{eqnarray}
f&\equiv& 
2\frac{a}{b^{2}} \left( 6\frac{a}{b^{2}} -\sqrt{2} \right) E^{2}_{(1)} \nonumber\\
&&+\left( -48\frac{a^{3}}{b^{6}}-4\sqrt{2}\frac{a^{2}}{b^{4}}+8\frac{a}{b^{2}} -\sqrt{2} \right) \frac{E_{(1)}L_{(1)}}{b} \nonumber\\
&&+\left( 4\frac{a^{2}}{b^{4}}-1 \right) \left( 12\frac{a^{2}}{b^{4}}+4\sqrt{2}\frac{a}{b^{2}} -1 \right) \frac{L^{2}_{(1)}}{b^{2}}
\end{eqnarray}
and
\begin{eqnarray}
g\equiv \frac{2aE_{(1)}}{b^{2}}+\frac{L_{(1)}}{b} \left( 1-\frac{4a^{2}}{b^{4}} \right).
\end{eqnarray}
It is easy to show that $f$ and $g$ are positive for $a/b^{2} \geq \sqrt{2}/2$. $L_{(4)}$ is negative since $L_{(3)}$ is positive. $V_{\mathrm{eff}(4)}''(0)$ is thus positive and the particle~(4) is bounded at the throat and it cannot exit the ergoregion.

\section{Concluding remarks \label{s-5}}

In the original Penrose process, a particle coming from infinity enters the ergoregion of a Kerr black hole and decays into a negative energy particle swallowed by the black hole and a positive energy particle escaping to infinity. Since the particle escaping to infinity has energy larger than the initial particle, the mechanism extracts energy from the black hole. In a collisional Penrose process, the difference is that the initial state is represented by two particles that collide in the ergoregion and create the negative and the positive energy particles.

The center-of-mass energy of the two in-going particles can be arbitrarily large in the vicinity of a Kerr black hole, but the efficiency of the energy extraction turns out to be only a little bit larger than 1~\cite{Bejger:2012yb}. The reason is that in the center of mass of the falling particles the radial component of the 4-momentum is large and negative near the event horizon, so only a small fraction of the collision energy can escape to infinity. Schnittman considered a scenario in which an initial particle bounces near a Kerr black hole and collides with a falling particle near the event horizon~\cite{Schnittman:2014zsa}. In such a case, the bounced particle does not have a large positive radial component of the 4-momentum since it collides near its turning point. The efficiency of the process can be larger by a few times than the case of two falling particles.

In this paper, we have investigated the collisional Penrose process in a wormhole spacetime. Our first result is that the process is indeed possible. In our discussion, we have employed the Teo wormhole spacetime, which represents the simplest rotating wormhole metric, and considered a head-on collision. However, we expect that similar results can be obtained in more theoretically motivated wormhole solutions. As for the particle collision, we have focused on two possible scenarios, namely the case in which the initial particles have vanishing angular momentum (flat potential case) and that in which the initial particles have negative angular momentum (deep potential case).

In the flat potential picture, the center-of-mass energy at the collision is small if the conserved energies of the initial particles are small. However, if the Teo wormhole rotates fast ($a\gg b^{2}$), 
the conserved energies of the particles produced in the collision can be very large, no matter the conserved energies of the initial particles, and the efficiency of the process to extract energy from the wormhole can be very high. This is in contrast to the Kerr black hole case. The upper bound on the total rest-mass of the particles created in the collision is determined by the center-of-mass energy. This implies that the rest-mass cannot be large if the conserved energies of the initial particles are not high in the flat potential case. In the deep potential case, even if the energies at infinity of the initial particle are low it is possible to get a high center-of-mass energy at the collision, with the possibility of an efficient extraction of energy from the wormhole and/or a large rest mass of the particle escaping to infinity. From these two scenarios, we learn that the center-of-mass energy of the particles at the collision is not very much related to the energy extraction from the wormhole, while it has a tight relation to the upper limit of the rest-mass of the particle escaping to infinity.

For simplicity, in this paper we have only considered a limited number of cases of particle collisions. It is thus likely that the actual upper bound of energy extraction from a Teo wormhole is larger than the upper bound found in this work. We have assumed that the sum of the radial component of the 4-momentum of the initial particles vanishes at the collisional point, but we could also consider a case in which it is positive there, with the result that a larger fraction can escape to infinity.

In the case of a black hole, the existence of the ergoregion and the possibility of extracting energy from the system have potentially important astrophysical implications, like the formation of powerful jets with the Blandford-Znajek mechanism~\cite{Blandford:1977ds}. It is plausible that a similar process is possible in the case of a rotating wormhole and, since the efficiency of the collisional Penrose process can be substantially higher than the same process around a black hole, the study of the Blandford-Znajek mechanism in a wormhole spacetime may be an interesting topic to explore in the future.

\section*{Acknowledgements}

We thank Tomohiro Harada, Umpei Miyamoto and Diego Rubiera-Garcia for useful comments and suggestions. This work was supported by the NSFC grant No.~11305038, the Shanghai Municipal Education Commission grant for Innovative Programs No.~14ZZ001, the Thousand Young Talents Program, and Fudan University.

\end{document}